# Modelling Families of Business Process Variants: A Decomposition Driven Method

Fredrik Milani, Marlon Dumas, Naved Ahmed, Raimundas Matulevičius

Institute of Computer Science, University of Tartu, Estonia, J. Liivi 2, 50409 Tartu, Estonia

`{milani, marlon.dumas, naved, rma}@ut.ee`

**Abstract.** Business processes usually do not exist as singular entities that can be managed in isolation, but rather as families of business process variants. When modelling such families of variants, analysts are confronted with the choice between modelling each variant separately, or modelling multiple or all variants in a single model. Modelling each variant separately leads to a proliferation of models that share common parts, resulting in redundancies and inconsistencies. Meanwhile, modelling all variants together leads to less but more complex models, thus hindering on comprehensibility. This paper introduces a method for modelling families of process variants that addresses this trade-off. The key tenet of the method is to alternate between steps of decomposition (breaking down processes into sub-processes) and deciding which parts should be modelled together and which ones should be modelled separately. We have applied the method to two case studies: one concerning the consolidation of existing process models, and another dealing with green-field process discovery. In both cases, the method produced fewer models with respect to the baseline and reduced duplicity by up to 50% without significant impact on complexity.

## 1 Introduction

Every organisation, be it non-profit, governmental or private, can be conceived as a system where value is created by means of processes [1]. Oftentimes, these processes do not exist as singular entities but rather as a family of variants that need to be collectively managed [2,3]. For example, an insurance company would typically perform the process for handling claims differently depending on whether it concerns a personal, vehicle or property claim [4]. Each of these processes for claims handling can be seen as variant of a generic claims handling process [5]. As such, processes with similar inputs and business goals, can be seen as variations of a single process in accordance with the definition provided in [5,6].

When it comes to modelling a family of process variants, one extreme approach is to model each variant separately. Such a *fragmented-model* approach [2] or a "*multi-model approach*" [5] creates redundancy and inconsistency. On the other hand, mod-elling multiple variants together in a *consolidated-model* approach [2] or "*single-model approach*"[5] leads to complex models that may prove difficult to understand, analyse and evolve. In addition to these comprehensibility and maintainability concerns, business drivers may come into play when determining whether multiple variants should be treated together or separately. Striking a trade-off between modelling each process variant separately versus collectively in a consolidated manner is still an open research question. In this context, our overarching research question is as follows.

"*How can a family of process variants be modelled?*"

(1) for consolidation of process models i.e. integrating a set of process models without changing the behaviour of business process, and

(2) for discovery of process models, i.e. green-field modelling of a business process.

The contribution of this paper is a decomposition driven method for modelling families of process variants. The core idea is to incrementally construct a decomposition of the family of process variants into sub-processes. At each level of the process model decomposition and for each sub-process, we determine if this sub-process should be modelled in a consolidated manner (one sub-process model for all variants or for multiple variants) or in a fragmented manner (one sub-process model per variant). This decision is taken based on two parameters: (*i*) the business drivers for the existence of a variation in the business process; and (*ii*) the degree of difference in the way the variants achieves the business goal of the process (syntactic drivers).

This article is an extension of a conference paper [7]. In the previous paper, we implemented the proposed method on a case study concerning the consolidation of existing process models. In this extended version, we validate the proposed method on a second case study where the goal is not to consolidate existing process models, but to capture a family of process variants from scratch. In this context, the proposed method is compared with a mainstream method for discovery of process models. Furthermore, the method is further refined with additional criteria for evaluating driver strength and similarity of variants.

The remainder of the paper is structured as follows. Section 2 introduces the conceptual foundation of our method. Section 3 describes the proposed method. Next, Section 4 introduces the case study method and the selection of case studies. Sections 5 present the application of the method to the case studies and Section 6 discusses the case study findings. Finally, Section 7 discusses related work while Section 8 draws conclusions and outlines future work.

## 2     Conceptual Foundation

The proposed method relies on two pillars: (*i*) a process decomposition method; and (*ii*) a decision framework for determining if two or more variants of a process/sub-process should be modelled together or separately. We present these two pillars in turn below.

### 2.1 Decomposition of Process Models

A number of methods for process decomposition exist [1,8,9]. Although these methods differ in terms of the nomenclature and specific definitions of the various levels of the process decomposition, they rely on a common set of core concepts that we summarise below.

A business process can be described at progressive levels of detail, starting from a top-level process, which we call the *main process* [9]. A main process is a process that does not belong to any larger process. The main process is decomposed into a number of *sub-processes* based on the concept of value chain introduced by Porter [8]. A sub-process is a process that is invoked by another (larger) process according to a call-and-return mechanism. Sub-processes are processes on their own and it can be further decomposed into sub-processes until such a level where a sub-process consists exclusively of atomic activities (called tasks) that do not warrant further decomposition.

Note that the above discussion refers to business processes, regardless of how they are represented. When modelling a business process, however, it is only natural to model each of its sub-processes separately. Accordingly, the hierarchy of processes derived via process decomposition is reflected in a corresponding hierarchy of process models representing the sub-processes in this decomposition.

### 2.2 Business and Syntactic Drivers

By applying incremental decomposition on a family of process variants, we reduce the problem of determining whether a given process should be modelled in a fragmented or consolidated manner, to that of deciding whether each of its sub-processes should be modelled in a fragmented or consolidated manner. To guide this decision, we propose a decision framework based on two classes of variation drivers. On the one hand, there may be business reasons for two or more variants to be treated as separate processes (or as a single one) and ergo to model these variants separately (or together). On the other hand, there may be differences in the way two or more variants achieve their business goals, which make it more convenient to model these variants separately rather than together or conversely. We refer to the first type of drivers as *business drivers* while the second type of drivers is called *syntactic drivers*.

Business drivers can range from externally dictated ones such as legislative requirements to internal choices an organisation has made such as organisational divisions due to mergers for example [10]. By categorising the many business reasons of process variations into *classes of variation drivers*, a reduction in complexity is achieved [11]. This enables working with a few classes of drivers rather than a multitude of possible root causes [12]. To this end, we use our previously presented framework [6], which is based on [1], for classification of business drivers.

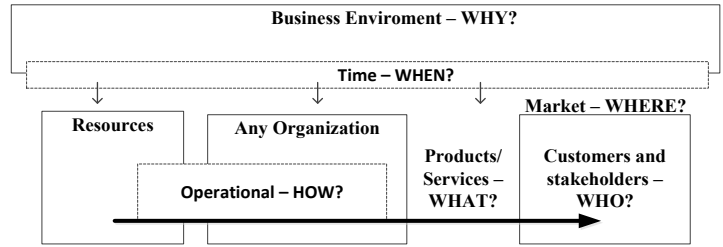

**Fig. 1.** Framework for classification of (business) variation drivers [6]

According to this framework (see Fig. 1), organisations operate within a context of external influences, to which they adapt their business processes. In this setting, organisations create an output by procuring resources in order to manufacture a product or a service (corresponding to *how* in Fig. 1). These products and services (*what*) are brought to a market (*where*) for customers (*who*) to consume. Organisations adapt their processes according to these aspects as well as their external environment such as tourist seasons (*when*). These adaptations lead to business process variations.

The key tenet of the framework is that business drivers for variations in business processes, based on their causes, can be classified as *operational (how), product (what), market (where), customer (who)* and *time (when)* drivers. This classification and the corresponding questions can be used to systematically elicit business variation drivers.

The second category of drivers that determine whether two variants should be modelled together or separately are syntactic drivers. If each variant were modelled separately, differences in the way variants achieve their business goals would naturally be reflected as differences between the models of each process variant. If these models differ in significant ways, it is more convenient to keep them separate, as consolidating them increases complexity and reduces comprehensibility. To capture this intuition, we assume that there exists a variant similarity function that given two variants, return a number between 0 and 1. Such similarity assessments can be made based on node matching, structural or behavioural similarity as discussed in [13].

Indeed, La Rosa *et.al.* [4] show empirically that the complexity of a consolidated model of two variants (measured by means of well-known complexity metrics such as size, density, structuredness and sequentiality) is inversely proportional to the similarity between the corresponding fragmented models, where similarity is measured by means of graph-edit distance. Hence, if we had a separate model of each variant, we could assess the complexity trade-off between merging them and keeping them separate, based on their graph-edit distance. However, existing approaches to measure process similarity require that (*i*) the models of the separate variants are available; and (*ii*) that they have been modelled using the same notation, at the same level of granularity and using the same modelling conventions and vocabulary. These assumptions are unrealistic in many practical scenarios where models of each variant might not be available to start with, and even if they were available, they would typically have been modelled by different teams and using different conventions.

When these assumptions do not hold, we propose to assess the similarity between variants of a process (or sub-process) by means of subjective judgment rather than a similarity measure. In other words, given two variants, we ask domain experts the question: How similar or different the separate models of these two variants would be if they were available?

In the following section, we operationalize the above concepts in the form of a method for consolidated modelling of families of process variants.

## 3    Method for Modelling Families of Process Variants

The proposed method for consolidated modelling of process variants follows the idea that decisions on whether to model two or more process variants together or separately, should not necessarily be taken at the level of the top-level process (main-process). Instead, such decisions should be postponed to each level of decomposition at the level of sub-processes. In other words, decisions on whether to model in a consolidated or fragmented manner should be interleaved with process decomposition steps. For instance, if two variants are extremely different, it is not optimal to start modelling them together as the number branching points will limit the readability of the model and most likely result in modelling the variants as separate models. Conversely, if two similar variants are modelled separately, the amount of duplication will be very high. As such, there will be an optimal point in the process hierarchy where decisions to model sub-process together or separately are to be taken. Depending on the particular process, that decision might happen at a higher or at a lower level of process hierarchy. Moreover, as the process is decomposed into sub-processes, a consolidated modelling approach for each sub-process should be the default option until it becomes clear that a fragmented approach is preferable from a business or syntactic perspective.

This observation, i.e. that decisions on whether to model in a consolidated or fragmented manner and that it varies from one process to another, was made from a case study on root causes of variations in business processes [6]. The method presented in this paper, is therefore derived from observations made in previous research [6].

These ideas are embodied in the following six steps as summarised below.

1. Model the main process – the purpose of this step is to elicit the main steps (sub-processes) of the business process in question.
2. Identify variation (business) drivers – in this step, the business drivers of variations are elicited so it becomes clear what drives the business process to have variability in the way it achieves its business goal.
3. Assess the relative strength of the variation drivers – in this step, the business drivers are analysed to determine which driver is the most important (strongest) driver of variations in the business process.
4. Identify the variants of each sub-process – at this point the actual existing variants of each sub-process previously elicited (in step 1) are identified and listed.

5. Perform similarity assessment of variants of each sub-process – at this stage of the method, the existing variants for each sub-process of the main process (or its corresponding sub-process at one level higher in the process hierarchy) are compared for the purpose of determining how similar or different they are from each other.
6. Construct the variation map – from the previous steps, the business drivers are present in the business process, the existing variants and their degree of similarity or difference are known. In this step, this information is used to determine if the variants of each sub-process should be modelled together or separately.

The steps are performed concretely with business stakeholders who have in-depth knowledge and understanding of the business process or parts of it. The roles that possess this knowledge may vary from organisation to organisation. In some organisations it is the business analyst, in others it might be subject matter experts or process owners. The actual list of participants (and roles) is naturally determined in discussion with those who have request the work.

### 3.1 Method Application

*Step 1 – Model the main process.*

In this step, the main process is modelled together with the domain experts and other relevant business stakeholders. The aim is both scope the business process in question and to identify the major (on high level) milestones in the business process. The level abstraction of the main process should be at such level as depicting the major $5\pm2$ steps (sub-processes) of the business process. One possible method for modelling the main process is introduced in [9] by Sharp and McDermott. In this method, start and end events are first identified and then the milestones are discovered. Alternative methods such as those introduced by Dumas et.al. [14], Harmon [15], and Rummler and Brache [1]. Although these methods vary slightly in how the main process elicited, they all provide the concepts necessary for modelling the main process and their further decomposition.

In Fig. 2, the main process of a governmental agency managing applications for constructions is shown. It starts with a plan being received and then it is registered, prepared, examined and finally approved. As such, the main process represents the major milestones of the business process.

*Step 2 – Identify variation (business) drivers.*

The second step is to elicit and to classify the business drivers for variations in the business process. Elicitation of variation drivers is achieved by using the framework presented in Section 2 (see Fig 1) together with two rounds of questions. In the first round, questions are asked about the existence of drivers in each of the categories of the framework (such as how many products/services the process produces or how many different customer segments the process serves). In the second round, each of these categories of drivers are further clarified and refined (such as how many sub-segments of customers are served in this process). For instance, in the example shown in Fig. 2, the first round of questions identified the existence of product drivers (new construction request or change to an existing construction). In the second round, the discussions clarified that new construction plans could be for either office or residential buildings. Concretely, this is achieved by means of a workshop or interview with business stakeholders. The output of this step is a list of all possible variation drivers for the business process.

*Step 3 – Assess the relative strength of variation drivers.*

Having identified the business drivers for the existence of variants in the business process under examination, a rating of importance (hereby called *strength*) is assigned to each driver. The strength of a variation driver reflects the level of investments needed to merge or standardize the process variants induced by the driver, as well as the level of management where decisions regarding such merging or standardization would be made. Importantly, the aim in this step is to rate the business importance of each variation driver, regardless of how much the variants differ from one another.

Variants induced by a "very strong" driver are fundamental to the business. For instance, if a company provides a service in two different markets, each with different regulations, the market driver is considered as "very strong". It would be very difficult (if even possible) to make changes in the variants across markets (such as standardising the two variants). Similarly, a product driver would be rated as "very strong" if a decision to substantially change the way one of the products is delivered would be seen as a change in the business model and would require a decision at the highest level of management. In other words, a very strong driver is such that its induced variants must be managed separately.

On the other hand, variants induced by a "strong" driver can in principle be merged or standardised to the point they can be managed together. However, their merging or standardisation requires significant investments and decisions from mid-to-upper management layers. For example, consider a company that has two different IT systems to support the same service due to historical or organisational reasons. A decision to merge or replace these two IT systems would require significant investment but would not affect the business model. Changes in variants induced by "strong" drivers are confined to individual business units and require decisions from the management of the business unit in question.

Variants generated by drivers rated as "somewhat strong" are considered to differ only at the level of minor details from a business perspective. In other words, whether these variants are managed together or separately is irrelevant from an upper management perspective. Change decisions on variants induced by a "somewhat strong" driver are taken at a low or mid-management tier.

Finally, a driver is rated as "not strong" if it is irrelevant to the business whether the variants are merged or kept separate, or in the latter case whether they are managed together or separately. For example, consider a company that provides the same service to two or more customer segments, such that differences in customer segments do not play a significant role in the way the service is sold and delivered. In this setting, the driver "customer segments" can be rated as "not strong".

The rating of drivers can be achieved via a workshop or interviews with domain experts, based on the questions outlined in Table 1. For a given driver, the first question, to which the answer is positive, determines driver's strength rating. If the answer to every question is negative, the variation driver is rated "not strong".

Only the strongest driver is taken as primary. If multiple drivers have equal strength, the one with fewer sub-categories is ranked higher. For instance, if an insurance company considers its products (individual travel insurance and business travel insurance) to be of equal strength as its customer drivers (northern, eastern, southern and western market segment), the product driver is ranked higher. The product driver has only two sub-categories whereas the customer driver has 4 sub-categories. In such cases, the driver with fewest sub-categories is ranked higher so as to reduce duplicity in the variation matrix.

In our running example, the primary variation driver was to be identified as the product driver with "new construction" and "change construction" plan. Having rated the relative strength of the variation drivers, this data is used to populate the first column of the variation matrix (see Fig. 2). The output of this step is a list of variation drivers for the process under examination together with their strength rating.

Table 1: Questions to help determine the strength of a driver.

| Rating | Question |
| --- | --- |
| Very Strong | Would a merger of the variants due to this particular variation driver be possible? Would a merger of these variants affect the business model or structure in such a fundamental way that it would require a decision from the highest level of management? |
| Strong | Would a merger of the variants (if desirable) require considerable investment, including noticeable re-organisation, and require decision from high level of management? |
| Somewhat Strong | Would a merger of the variants (if desirable) require some investment, include some re-organisation noticeable to the concerned business unit only, and require decision from mid-level management? |
| Not Strong | None of the above. |

*Step 4 – Identify the variants of each sub-process.*

In the fourth step, existing variants for each sub-process identified in step 1 and for each variation driver are identified. This is concretely done by asking the business stakeholders for each sub-process, existing variants per business driver and adding them, one by one, to the variation matrix. The variants are therefore captured in a textual way by their name. The output is a variation matrix (see Fig. 2) wherein the rows correspond to business drivers (qualified by their relative strength) and the columns correspond to the sub-processes identified in step 1. A cell in this matrix lists the variants of a given sub-process (if any) induced by a given driver. For conven-ience the drivers are listed in descending order of strength (see Fig. 2). For instance, in our running example, there are three different variants for examining a plan (examine NCP off, examine NCP res and examine CCP).

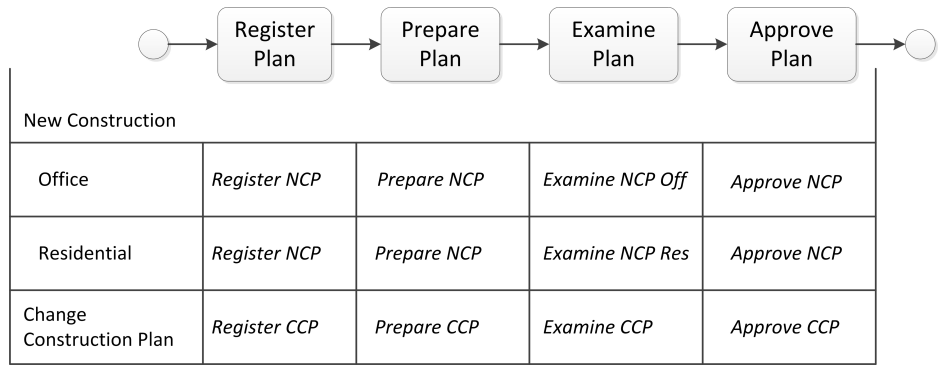

**Fig. 2.** Variation matrix

*Step 5 – Perform similarity assessment of variants of each sub-process.*

In this step, we perform a similarity assessment for each subset of variants of each sub-process identified before. As discussed in Section 2, for this similarity assessment we do not assume that detailed models of each sub-process are available for comparison.[1] Accordingly, we employ a 4-point scale for similarity judgements extensively used in the field of similarity assessment [16]: (1) very similar, (2) similar, (3) somewhat similar, and (4) not similar (identical variants are marked as identical and not subjected to similarity assessments).

This step can be implemented by interviewing the domain experts, asking them – given the identified variants of each sub-process – if the variants of the sub-process are likely to lead to models that are identical, very similar, similar, somewhat similar or not similar (see Table 2). The output of this step is an annotated variation matrix, where is set of variants of a sub-process is annotated with the result of their similarity assessment.

For instance, in Fig. 2 there are two variants of register plan (register NCP and register CCP). The business stakeholders are asked about the similarity of these two variants. Colour codes or any other method of choice can be used to distinguish the similarity of the sub-process variants. If more variants are available, such as in the case of "examine plan" in Fig. 2, the same procedure is repeated but beginning with variants within one variation driver first. For instance, the similarity of "examine NCP Off" is compared with "examine NCP Res". Then, they are assessed as compared to "examine CCP".

---

[1] Although if such models are available they can naturally be used.

Table 2: Guideline for Subjective Assessment of Similarity

| Similarity Assessment | Similarity of two variants |
|---|---|
| Identical | There is no perceivable difference. |
| Very Similar | Differences can be perceived but they are not significant. |
| Similar | There are clear similarities through out the process. |
| Somewhat Similar | There are some isolated parts of the process that are perceivably similar. |
| Not Similar | There are in essence no perceivable similarities |

*Step 6 – Construct the variation map.*

From the previous steps, we know the strength of the business drivers and the degree of similarity between the variants of each sub-process induced by a driver. This information is used to assess the trade-off of modelling the variants in a consolidated versus fragmented manner. In making a decision to consolidate or fragment, the analyst will use the decision matrix depicted in Fig.3.

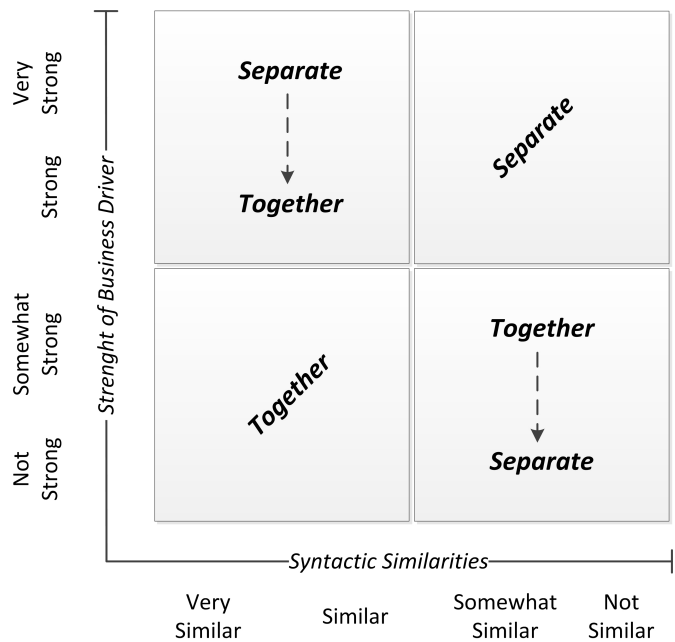

**Fig. 3.** Decision matrix for modelling variants separately or together

If the variants are very similar and there are no strong business drivers for variation (not strong or somewhat strong i.e. no significant business impact), then naturally the variants are modelled together. Conversely, if there are strong business drivers (strong or very strong i.e. they have business impact) and the variants are syntactically different (somewhat similar or not similar) they are modelled separately. If variants are similar and have strong business drivers, they are modelled together or separately depending on the current level in the process decomposition. The levels of decomposition are either high or low. By high level of decomposition, we refer to level 3 (level 1 and 2 refer to Business Model and the main process accordingly) of the value creation system hierarchy introduced by Rummler and Brache [1]. Using the same process architecture, low levels of decomposition refer to levels 4 and 5 (lowest levels of decomposition).

At levels close to the main process (high levels), sub-process variants falling in this quadrant are modelled separately because the business driver for separating the variants prevails. If the business driver is strong, it pre-supposes that the variants have different process owners and stakeholders and therefore the modelling effort has to be done separately for each variant.

At lower levels of process decomposition, the business driver for modelling two variants separately weakens down and the incentive for sharing the modelling effort for variants increases. Therefore for sub-processes at lower levels of decomposition, the syntactic driver prevails, i.e. if these processes are similar, they are modelled together as a consolidated sub-process. Conversely, in the lower right quadrant, variants of sub-processes at a high level of decomposition are modelled together, since these variants fall under the same ownership area and thus it makes sense to conduct a joint modelling effort for them. However, at the lower levels of decomposition, if two sub-process variants are not similar, the analysts can choose to model them separately.

The output of this step is a variation map (see Fig. 4). A variation map is a process model where there are only tasks and XOR splits, representing the points where multiple variants will be separated. In constructing the variation map, only the allowed combinations are modelled using gateways. As such, the variation map shows the variants of each sub-process that ought to be modelled separately. The variation map contains one decision gateway per subset of variants of a sub-process that needs to be modelled separately. If a sub-process does not have variants, it is not preceded by a gateway.

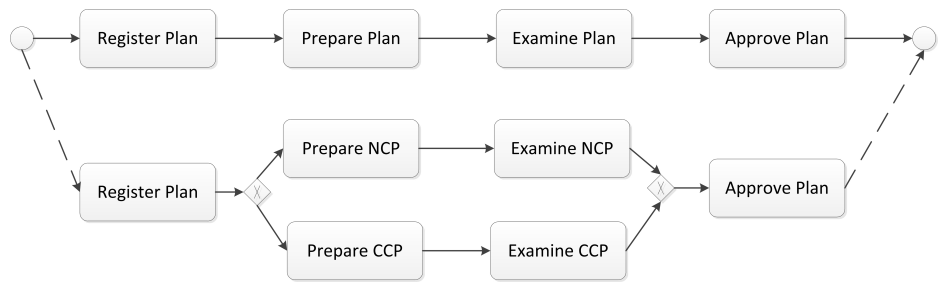

**Fig. 4.** Variation map

For instance, having performed the similarity assessment based on the variation matrix in Fig. 2, it has become known that variants "Examine NCP Off" and "examine NPC Res" are very similar to each other but different from "Examine CCP". As such, as we are at a high level of decomposition, the variants for NPC are modelled together and CCP is modelled separately from NCP as according to the decision ma-trix in Fig. 3. Furthermore, when constructing the variation map, constraints between variants of pre- and succeeding sub-processes are considered. For instance, as depicted in Fig. 4, only "examine NCP" can follow "prepare NCP", i.e. it is not possible to execute "examine CCP" after "prepare NCP". The same procedure is then repeated for each level of decomposition.

Having constructed the variation map for the first level of process decomposition, we consider each of the sub-process variants in the variation map in turn. Each sub-process variant is then decomposed into a lower-level process model and steps 2-4 are repeated at that level.

## 4      Case Study Methodology

A case study is defined as an empirical method with the purpose of investigating a certain reality within its real-life context [17], particularly when the boundaries between what is studied and its context are not clear [18]. Case studies are often used for exploratory purposes, but they are also suitable for testing a hypothesis in a confirmatory study [17,19] or to evaluate a method within the software and systems engineering domain [20]. These features make the case study method applicable as an instrument to validate our proposed method.

### 4.1      Research Questions

Yin [18] argues for the necessity of defining a research question when designing a case study. Our overarching research question and its two sub-questions are:
  *RQ: How can a family of process variants be modelled?*
  *RQ 1: How can a family of process variants be consolidated in a manner that results in the usage of fewer activities and sub-process models?*
  *RQ 2: How can a family of process variants be discovered in a manner that results in the usage of fewer activities and sub-process models?*

Furthermore, Yin [18] states that there is a need for developing a hypothesis. The purpose of our method is to manage variants process models that have less redundancy than a collection of fragmented models. Thus, our hypothesis is that *"when our method is applied on a family of process variants, then the same set of business processes can be represented using fewer activities and sub-process models than if the same was done using a fragmented approach."* We do not expect (our alternative hypothesis) that *"when applying our method, the size of the family of process variants is the same or larger in terms of total number of activities and sub-process models compared to a fragmented approach."*

These research questions are relevant given that reducing the number of activities, in particular duplicates, and sub-processes, lead to better comprehensibility of the process models [21], less duplicity and stronger linking of related sub-processes. This in turn, will reduce maintenance efforts and will also facilitate the analysis, comparison and implementation of process variants in a common IT system[5].

Given the above research question, we sought case studies where families of process variants needed to be managed collectively. Naturally, we also sought case studies where we could engage with domain experts, as our method heavily relies on their input. Finally, we looked for case studies that would allow us to address both research sub-questions and that were representative of different modelling purposes, industry sectors, level of IS maturity and transaction volumes. Below we present the selected case studies and the case study design.

### 4.2 Case Study Settings

The organisational setting of our first case is the foreign exchange (FX) and money market (MM) operations of a mid-sized European bank. FX covers financial products related to the trade of international currencies. MM covers trade in short-term loans and deposits of financial funds between various institutions. Currently, the bank employs a legacy system for managing these products (families of process variants) but want to replace it with an off-the-shelf system. For this purpose, the bank needs to elicit requirements for all variants, which primarily come from the corresponding business processes rather than from the current IS structure. The business processes had, several years before this case study been modelled as separate process models (4 main flat flowcharts with more than 200 activity nodes) by a team of consultants. The existing models were flat (no decomposition had been made). Three of these models were for the variants of the process related to trading FX and MM with interbank counterparts and one for non-interbank clients who do not have an account with the bank. The bank aims at consolidating these process models prior to requirements elicitation.

The second case is from a Genome Centre – a small-sized semi-publicly-funded organization engaged in research and development related to DNA sequencing and analysis. The centre performs DNA sequencing both as part of their own research and as service provided to other academic and corporate institutions. At present, the centre manages their processes manually. However, they intend to implement a Laboratory Information Management System (LIMS) that would allow them to better plan, perform and monitor their sequencing projects. At present, the workflows for genome sequencing are documented as textual documents in more than 40 different protocols, each one describing a specific variant. In order to elicit requirements for the future LIMS, the centre decided to model their sequencing process across all its variants.

We note that the context of the first case study matches RQ1 (consolidation of existing models of process variants) while the second case study matches RQ2 (from-scratch modelling of a family of process variants). Also, the case studies differ in purpose, industry sector, level of IS maturity, transaction volumes and level of modelling experience of the domain experts. In the banking case study, the purpose was to combine requirements of different variants for the purpose of evaluation of options for replacing multiple IS with a single one, whereas in the second case, the purpose was to elicit requirements for the purpose of building a completely new IS. With respect to industry sector, level of IS maturity and transaction volumes, the two case studies are very distinct. The first case study is from the banking industry and in-volves large number of transactions mostly managed by several highly integrated and automated IS. The second case study, on the other hand has low volumes of transactions and involve a high level of manual processing with almost no IS support. Regarding the context, in the banking case, we began with a set of process models that needed to be consolidated (bottom-up approach), whereas in the DNA case, the process models were discovered (top-down approach). Finally, regarding modelling experience, the experts of the banking case study had at least 5 years of experience with process models as opposed to the domain experts of the DNA case that had not worked with process models at all.

### 4.3 Case Study Design

The design of both case studies (see Fig. 5) consists of eight steps, out of which the first six correspond to the steps in the method introduced in Section 3, while the seventh step corresponds to working with the process models (constructing the consolidated (sub-)process models for our first case study and the modelling of the business processes for the second case study). The final step consists of verifying the process models that have been produced with the domain experts.

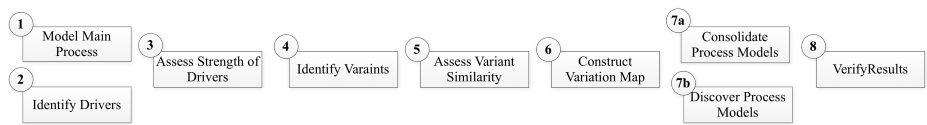

**Fig. 5.** Case Study Design

There are slight differences in the execution of the case studies. The reasons are that (1) the business processes of the banking case study had already been modelled whereas in the DNA case study, there were no process models. Furthermore (2) the domain experts for the banking case study had all at least 5 years of experience with process management whereas the domain experts of the DNA case study, did not have any experience with process models.

One difference is that we began the first case study with modelling the main process, but in the second case study, we found it more meaningful to start with identifying the variation drivers first and then model the main process.

Secondly, in order to examine the effectiveness of our method for the second case study, we need a baseline scenario or baseline process models (comparable to the input process models in the first case study) to which we could compare our results. For this purpose, we modelled the DNA process models according to our method and according to guidelines presented by Sharp & McDermott [9]. We choose it because (1) this approach is widely used and recommended in practice [22] and (2) the approach explicitly takes into consideration variations in business processes and argues for a more fragmented approach when modelling process variants.

## 5  Case Study Execution

This section describes the execution of the two case studies. First we describe the first case study (FX&MM), followed by the execution of the second case study (DNA) using our method and the baseline method.

### 5.1  Execution of FX & MM Case Study

The banking case study was conducted as described in Section 4.4. We applied our method in a four-hour workshop with five domain experts, led by the first author of this paper. In addition, two stakeholders from IT support were available for questions and clarifications. The workshop resulted in a variation map of the business processes. The first five steps were conducted in one workshop and in total took 4 hours. The first step (modelling the main process) took less than an hour and the elicitation and classification of drivers also took less than one hour. The similarity assessment, with the aid of the variation matrix, took around two hours. Afterwards, the variation map was modelled, which together with its verification, took three hours. The actual consolidation of process models took roughly 80 man-hours to complete. Finally, the verification of the consolidated process models were done by the domain experts in a series of 8 workshops, each taking 1.5 hours on average to conduct.

*Step 1 - Model the main process of FX&MM trades.*

In the first step, we modelled the main process for managing FX&MM trades (see Fig. 6). We started by asking what initiates the process and then, through a series of questions, modelled each step of the process until the end. This step resulted in a model of the main process for FX&MM products (see Fig. 6).

The main process is initiated once an order is received. The first task is to "*Register Trade*" meaning entering the trade in the IS. The next task is "*Approve Trade*". Then, "*Confirm Trade*" takes place when the bank sends a confirmation of the trade details to the counterpart. Once the counterpart "*Match Trade*", i.e. agrees to the trade data, "*Settle Trade*" takes place (transfer of payment). The final task is "*Book Trade*" which is when the trade is booked in the accounting systems.

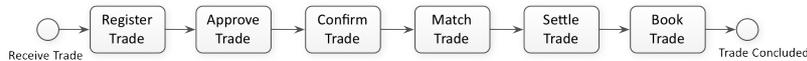

**Fig. 6.** Main process for managing FX & MM trades

*Step 2 – Identify the variation drivers.*

The second step (see Fig. 5) was to identify variation drivers of the process. We started by introducing the concept of variation drivers and the framework (see Section 2) for their classification. We then gave some examples of variation drivers and asked the domain experts if their business processes have occurrences of such variation drivers.

We observed that product and customer driven variations existed. The product driven variations were FX, MM and NDF (non-deliverable forward i.e. trading in restricted currencies). The customer driven variations were identified as Bank (other banks), Corporate (companies), Private (individuals) and Site (belonging to branches) clients. Furthermore, the corporate clients had a customer driven variation of type account (having an account agreement with the bank) or cash (do not have an account with the bank) client type.

*Step 3 – Assess the relative strength of the variation drivers.*

As input from the previous step, i.e. having the variation drivers identified, we continued with determining their relative strength. Through discussions we understood that the product drivers were the strongest. It also became clear that FX & MM were similar enough to be treated as one. However, NDF is separate and on its own.

*Step 4 – Identify the variants of each sub-process of the main process.*

With the input from the previous steps, we could populate the variation matrix (see Fig. 7). First, we used the variation drivers and their relative strength to populate the first column of the variation matrix. Then, for each sub-process of the main process, such as *"match trade"*, we ask the domain experts, how is this process performed? For instance, for an FX trade done with another bank, the ways to match the trades are either Intellimatch (in-house trade by trade matching) or CLS (a centralised intra-bank platform). We thus enter these two variants in the matrix under sub-process *"match trade"* and for customer type *"bank"* (see Fig. 7). Note that in this case, the same solution (such as CLS) is used in "match trade" and "settle trade". As CLS is a centralised intra-bank platform, it has several functions and therefore used in two or more sub-processes. However, the use differs i.e. how CLS is used in "match trade" differs from its use in "settle trade". As such, in this case study, although the variants may bear the same name, they differ as they are situated under different sub-processes of the main process.

*Step 5 – Perform similarity assessment of variants for each sub-process of the main process.*

We performed the similarity assessment by visiting each cell of the variation matrix in turn. For example, the variation matrix shows that corporate and site clients have the same variants for matching a trade. We first asked the domain experts to identify identical variants if there were any and then to grade the level of similarity of remaining variants from 1 (very similar) to 4 (not similar). The results showed that all swift trades are very similar. The same applied to platform, online and paper. Furthermore, the domain experts assessed that swift, platform, online and paper are similar to each other as the process is basically the same but the tool used to match the trades differs depending on customer type. For instance, the process by which a trade by swift or paper is similar but differ in what medium is used (swift or paper). We also observed that matching in bulk (when several trades are matched at once) is very different compared to matching by SWIFT, platform, online and paper. As mentioned before, these variants are only compared to other variants under the "match trade" sub-process.

Having established the degree of similarity among the corporate, private and site clients, we continued asking about similarities between CLS and Intellimatch for when the counterpart is a bank. These differed significantly compared to how trades are matched for non-bank counterparts. This step resulted in identifying two main variants for matching when the counterpart is a bank (i.e. Intellimatch and CLS) and two main variants when trading with non-bank counterparts (i.e. when the matching is done in bulk versus single-trade match).

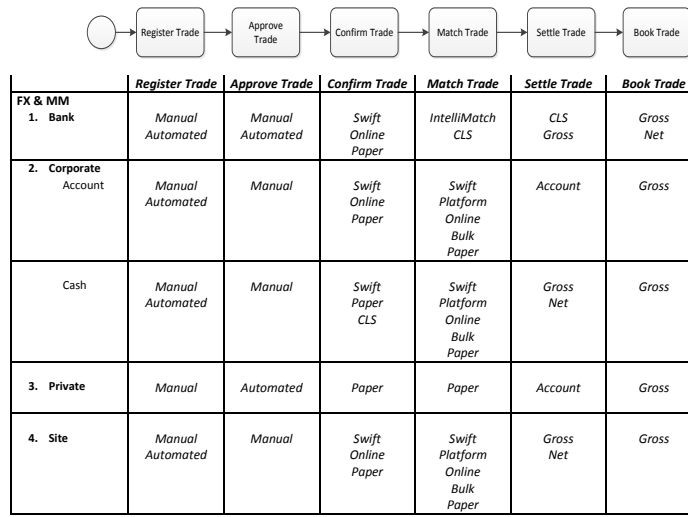

|  | *Register Trade* | *Approve Trade* | *Confirm Trade* | *Match Trade* | *Settle Trade* | *Book Trade* |
|---|---|---|---|---|---|---|
| **FX & MM** <br> **1. Bank** | *Manual* <br> *Automated* | *Manual* <br> *Automated* | *Swift* <br> *Online* <br> *Paper* | *IntelliMatch* <br> *CLS* | *CLS* <br> *Gross* | *Gross* <br> *Net* |
| **2. Corporate** <br> Account | *Manual* <br> *Automated* | *Manual* | *Swift* <br> *Online* <br> *Paper* | *Swift* <br> *Platform* <br> *Online* <br> *Bulk* <br> *Paper* | *Account* | *Gross* |
| Cash | *Manual* <br> *Automated* | *Manual* | *Swift* <br> *Paper* <br> *CLS* | *Swift* <br> *Platform* <br> *Online* <br> *Bulk* <br> *Paper* | *Gross* <br> *Net* | *Gross* |
| **3. Private** | *Manual* | *Automated* | *Paper* | *Paper* | *Account* | *Gross* |
| **4. Site** | *Manual* <br> *Automated* | *Manual* | *Swift* <br> *Online* <br> *Paper* | *Swift* <br> *Platform* <br> *Online* <br> *Bulk* <br> *Paper* | *Gross* <br> *Net* | *Gross* |

**Fig. 7.** Populated Variation Matrix (NDF excluded due to space limitation)

*Step 6 – Construct the variation map.*

As input from step 4, we have the variants for each sub-process of the main process and are able to map them in a variation map (see Fig. 8). For instance, we had two variants of *"Register Trade"* (manual and automated). These sub-processes did not have a strong business driver and were similar. Referring to the decision framework (see Fig. 3), we modelled them together. Conversely, there are two variants of *"Settle Trade"* for bank clients in the variation matrix in Fig. 7 (CLS and gross). These were assessed to have a strong variation driver and also to be not similar. As such, in accordance with the decision framework (Fig. 3), they are modelled separately. After having continued in the same manner for all sub-processes, the variation map for each sub-process was constructed as sub-process is depicted in Fig 8.

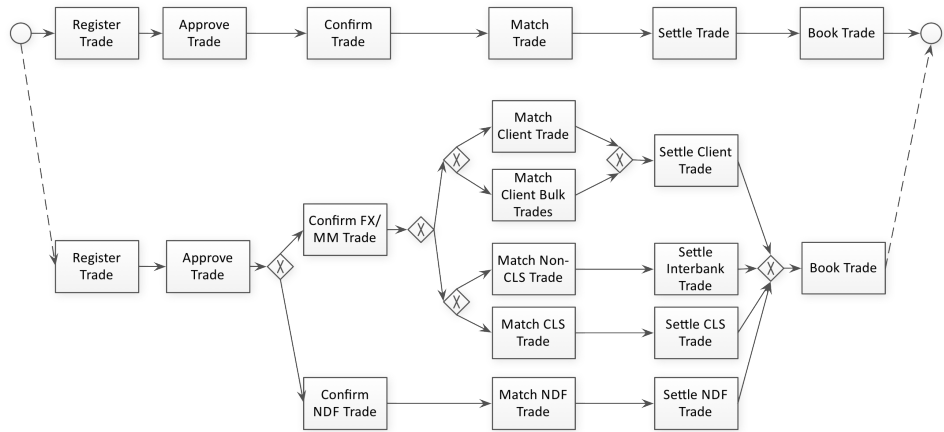

**Fig. 8.** Variation map for FX&MM main process

*Step 7a - Consolidation of Input Process Models.*

The original process models had been modelled as flat end-to-end process models. As a first step, we divided these models into sub-processes in accordance with the decomposition identified in step 3. This gave us four hierarchical process models: one for FX traded gross, one for FX traded via CLS, one for MM, and one for corporate clients. In addition to these four process models, there were two additional processes described as text, one for NDF and one for bulk matching, which we modelled diagrammatically as part of the consolidation effort.

For each task of the main process, we compared and consolidated them in accordance with the variation map. We sought clarification from the domain experts and IT stakeholders when needed. The input process models had not been regularly updated with changes in the business processes during the past 3 years and therefore we observed minor discrepancies. We updated the consolidated process models accordingly.

*Step 8 - Verification of Consolidated Process Models.*

Once the process models had been consolidated, domain experts verified them (in detail) in a series of 8 workshops. The coordinating domain expert made adjustments to the consolidated models during these workshops. After all workshops, the domain experts were asked about the usefulness of the models in terms of comprehensibility and if they will use the models for evaluating off-the-shelf systems. They stated that the consolidated models are easier to understand (compared to the input process models). The consolidated process models were used as standard evaluation criteria for finding suitable replacement systems for their FX/MM business processes. Consolidated process models facilitate designers to analyse, compare and implement IT systems that are to support the process variants [5]. As such, the direct link of our method to information systems in this particular case study is its use in finding suitable replacement for an off-the-shelf system. The consolidated models were used for eval-uating one product to support these FX/MM processes and to compare it to alternative products.

### 5.2 Execution of DNA Case Study using the Decomposition Driven Method

For the DNA case study, the first and third author of this paper conducted three initial meetings with the head of the sequencing lab and two domain experts. During these meetings, information needed for constructing the variation map was gathered (steps 1-6). These steps took in total about 5 hours. We then proceeded with modelling the processes in details (step 7b) followed by verification. The modelling effort itself (by the two authors) amounted to circa 360 person-hours followed by 40 person-hours of verification by the domain experts.

*Step 1 – Identify variation drivers.*
For the DNA case study, we did not have any process models and the domain experts had no experience with process modelling. We therefore decided to conduct the identification of variation drivers first. We started by introducing the concept of variation drivers and their classification. We then proceeded to ask the "W" questions, as discussed in Section 2, in order to uncover variation drivers. We identified the existence of one product driver and one operational driver. The product driver is the co-existence of two distinct services: DNA sequencing (determining the order of nucleotides of sample containing DNA) and array analysis (analysing the genetic makeup of a DNA that determines specific traits). The operational driver relates to which machine is used for sequencing or analysis.

*Step 2 – Model the Main Process of the DNA Sequencing Process.*
The second step was to model the main process. We asked what triggers the process, which milestones the process goes through and what value each step produces. This led to the main process depicted in Fig. 9. The process is triggered when there is an agreement with a customer to sequence some samples. Then, the project data are registered followed by the samples being prepared. Once the samples are prepared, they are processed, meaning that they are sequenced using the sequencing and analysis machines. In the final step, data is extracted and delivered to the customer.

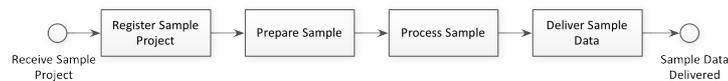

**Fig. 9.** Main process for Sequencing DNA Samples.

*Step 3 – Determine the relative strength of the variation drivers.*
We used the questions in Table 1 to assess the strength of the variation drivers. This resulted in identifying the product driver (sequencing versus array analysis) as the strongest followed by the operational driver (HiSeq, MiSeq or Array Machine).

*Step 4 – Identify variants of each sub-process.*

We performed step 4 in the same manner as we did the *banking* case study. This resulted in identifying, for sample preparation using HiSeq, two variants namely "Prepare TrueSeq Sample" and "Prepare Nextera Sample" (see Fig 10). We can also see from the variation matrix, that the same variants exist for preparing a sample when using MiSeq machine for sequencing. Similarly, we note that there are three different variants for Array Sequencing, one for processing DNA samples, one for RNA and one for Methylation.

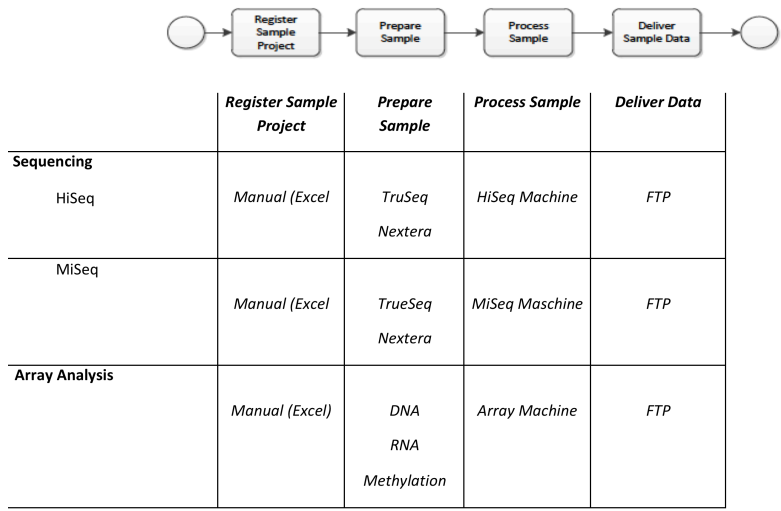

|  | *Register Sample Project* | *Prepare Sample* | *Process Sample* | *Deliver Data* |
|---|---|---|---|---|
| **Sequencing** | | | | |
| HiSeq | *Manual (Excel* | *TrueSeq* <br> *Nextera* | *HiSeq Machine* | *FTP* |
| MiSeq | *Manual (Excel* | *TrueSeq* <br> *Nextera* | *MiSeq Maschine* | *FTP* |
| **Array Analysis** | | | | |
| | *Manual (Excel)* | *DNA* <br> *RNA* <br> *Methylation* | *Array Machine* | *FTP* |

**Fig. 10.** Populated Variation Matrix for DNA Sequencing

*Step 5 - Perform similarity assessment of the variants for each sub-process.*

During the workshop, we used different colours of whiteboard pens to annotate the variants of each sub-process that were very similar, similar and so forth, leading to an annotated variation matrix.

*Step 6 – Construct the variation map.*

With the input from the previous step, we can proceed with deciding which variants are to be modelled together and which are to be modelled separately, resulting in the variation map depicted in Fig 11.

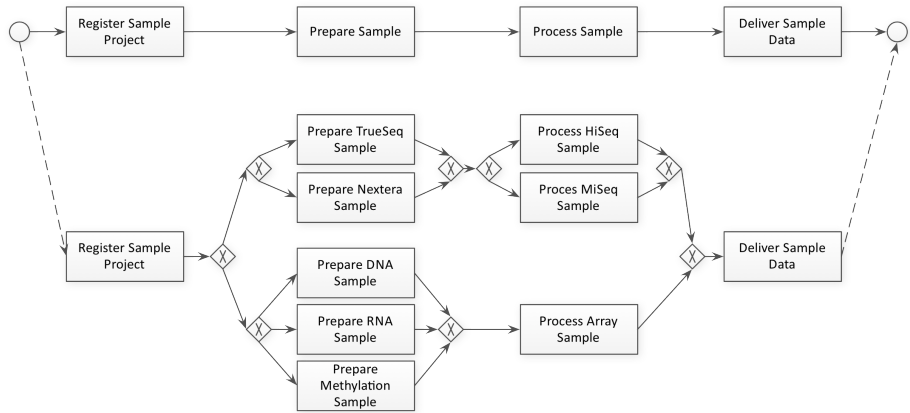

**Fig. 11.** Variation map for DNA sequencing main process

*Step 7b - Discovery of Process Models.*

In this case study, there were no process models to begin with. There were, however "protocols" that explain on a very detailed procedural level, the steps to be performed for each specific case (variant). The lab has a little over 40 different "protocols" where the shortest is 20 pages and the longest is around 200 pages (about 4000 pages in total). These "protocols" include processes that the DNA sequencing lab does not employ. On the other hand, the protocols cover only two sub-processes of the main process (Prepare Sample and Process Sample). The other sub-processes had not been previously documented.

For each sub-process of the main process, we modelled them in more detail in accordance with the variation map from step 6. The process models were discovered and modelled from the protocols and additional input from domain experts gathered via hour-long weekly meetings for a period of 6 weeks.

*Step 8 - Verification of results by domain experts.*

Once all processes had been modelled, we organized a hand-over meeting with the domain experts and presented the structure of the models. Following this meeting, the models were handed on to the domain experts for detailed examination. As not all domain experts work with the same processes or have the same responsibilities, they divided the sub-processes in accordance with their area of expertise and responsibility. Thereafter, we had weekly meetings with the domain experts for 6 weeks. At each meeting, we verified a subset of process models they had examined. After the verification had been completed, we continued to engage with the domain experts in weekly meetings to elicit requirements from the process models for their future LIMS. As such, the process models were used as the primary source for eliciting requirements for their new information system. In fact, the first prototype of their information system was developed on the basis of the requirements elicited form the process models discovered using our method.

During these requirements elicitation meetings, we observed on multiple occasions that variants of a sub-process, which had been modelled together, had in fact visible differences when considering the objects manipulated by the tasks in the process. While anecdotal, this observation suggests that variants that are procedurally very similar might differ significantly at the level of (data) objects. This observation deserves a separate study as discussed in Section 8.

### 5.3    Execution of DNA Case Study using the Baseline Method

As mentioned in Section 4.4, we modelled the same set of DNA sequencing processes according to the approach outlined by Sharp and McDermott [9], so as to have a baseline for comparison. Sharp & McDermott's method (henceforth S&M) consists of the following steps:

Step 1: Identify the start events (called triggers) and end events (results) of the process.
Step 2: Identify major components based on milestones of the process (sub-processes).
Step 3: Identify the variants (cases) of the process.
Step 4: Identify internal and external stakeholders (participating organizations in the process).
Step 5: Identify for each participating organization, the individual actors their main responsibilities.
Step 6: Identify systems and data objects (supporting mechanisms) of the process.
Step 7: Conduct workshops where all tasks are listed and then sorted in order to create a flat process model.
Step 8: Identify the logical breakpoints of the flat process model and cluster activities within these points together as a sub-process.

The S&M method adopts a fragmented approach when modelling families of process variants. The method advocates for keeping variants in separate process models, arguing that design-time variation points should not be captured in a process model because these decisions have already been made prior to, and not during the execution of the process. However, if two variants are very similar, the method concedes that they can be modelled together, although this is not the preferred solution. Concretely, in case multiple variants have been identified, the method suggests to start by modelling one variant completely – for instance the most common one. This first variant is modelled flat. Next, the second variant (case) is taken and compared to the already modelled variant. If the variants are very similar, they can be modelled together. Note that the first five steps are conducted only once and step 7 and 8 are repeated for each additional variant. As such, for each additional variant that is different from the first process model, step 7 and 8 are repeated.

In order to minimize learning effects, we applied the S&M method in parallel with our method. The first two steps of S&M concern modelling the main process as part of the purpose of framing the project. Accordingly, while performing step 1 of our decomposition-driven method (modelling the main process), we gathered information needed for performing the first two steps of S&M. Similarly, step 2 (identify drivers) and step 4 (identify variants) of the decomposition-driven method correspond to steps 3 to 6 in S&M method, and thus these steps were done in parallel for both methods.

Steps 7 and 8 of the S&M method are the steps were the models are produced. Two of the authors of this paper began by modelling the most common variant, first as a flat process model, followed by the identification of logical breakpoints and extraction of sub-processes. Then we proceeded with the next variant by comparing it with the first already modelled variant. If they were very similar, we modelled them together. This procedure we repeated until all variants of the main process had been covered. Steps 7 and 8 of the S&M method were performed in parallel with Step 7b of the decomposition-driven method (see above).

It should be noted that the S&M method does not provide guidance as how to manage sub-processes that are shared by several variants. We chose to apply refactoring for this purpose, i.e. if several variants shared a sub-process, we modelled it only once.

## 6    Findings

### 6.1    Results from the Banking Case Study

As mentioned earlier, in the banking case study the original (input) process models had been modelled flat (no decomposition). In order to make them comparable with the models produced after consolidation, we split each of the flat process models into sub-processes following the same sub-process structure that resulted from the consolidation. In this way, the input process models and the consolidated ones are comparable in terms of hierarchical structure, although they differ in amount of duplication.

The input process models did not include NDF and bulk matching. These processes had only been partially documented in textual format prior to the consolidation. During the consolidation effort, these two processes were modelled as well. However, to make the input and the consolidated process models comparable, we do not take into account NDF and bulk matching in any of the statistics given below.

The input process models contain 35 sub-process models and 210 activity nodes (not counting gateways and artefacts such as data objects or data stores). Out of these, 75 activity nodes were duplicate occurrences (an activity occurring N times across all sub-process models counts as N duplicate occurrences). Thus, it can be said that the duplication rate in the input models is 36 %. Note that the 35 sub-process models in the input were distinct models, although some of them had strong similarities.

The consolidated models contain 17 sub-process models and 149 activity nodes of which 22 are duplicate occurrences, corresponding to 15 % duplication. Thus the consolidated models contain 30% less activity nodes, half of the sub-process models and half of the duplication rate relative to the original model. These observations (summarised in Table 3) support the hypothesis of the case study.

**Table 3.** Size metrics before and after consolidation

| Variable | Input | Consolidated |
|---|---|---|
| Main Process Models | 4 | 1 |
| Sub-Process Models | 35 | 17 |
| Activity Nodes | 210 | 149 |
| Duplicate Activity Occurrences | 75 | 22 |
| Duplication rate | 36 % | 15 % |
| CNC | 1,25 | 1,33 |

One can expect that the complexity of the process models will increase during consolidation since additional gateways are introduced to capture differences between multiple variants of a sub-process model. For instance, the consolidation naturally affected the complexity measure. For instance, there were four separate sections of the flat process models corresponding to "register trade". The input process model for corporate clients (trading both FX and MM) was the least complex one with a CNC of 0.8. For interbank trading of FX (both gross and CLS), the corresponding CNC was 1.09. For the interbank trading of MM, it was 1.17. The combined complexity of the input "register" process models were 1.07. The consolidated sub-process (only one as the variants were similar and lacking strong variation driver), have a CNC of 1.11. This trade-off between reduction in duplication and increase in complexity has been observed for example in [4].

To measure the impact of consolidation on complexity, we use the Coefficient of Network Complexity (CNC) metric. CNC is the ratio between the number of arcs and the number of nodes. This simple metric has been put forward as suitable for assessing the complexity of process models [23]. The input process models had a total of 350 arcs and 280 nodes (210 activity nodes and 60 gateways and start/end events). This gives a CNC of 1.25. The consolidated process models consist of 320 arcs and 240 nodes (149 activity nodes and 81 gateways and start/end events) giving a CNC of 1.33. Thus, there is a marginal increase in complexity as a result of consolidation. This should be contrasted to the significant reduction in size and duplication.

The input process models had four main processes, one for corporate clients trading both FX and MM, two for interbank trading of FX via gross or CLS and finally one for interbank trading of MM (both gross and CLS). Therefore, there was no distinct driver behind the segregation of the process models. In one case it was based on customer type (corporate versus interbank) regardless of product. In another case it was based on product (FX versus MM) and a third one was based on how the trades were settled (gross or CLS). In contrast to this, the consolidated models had one common main process, where the business drivers for variations are expressed at each sub-process of the main process. For instance, for "confirm trade" the driver is based on product (FX/MM versus NDF), and for "match trade" it is based on customer (corporate versus interbank). As such, the consolidated set of process models were also a restructuring of how the business process is captured. The structure of the main process changed from four flat input main processes to one that encompasses all four by expressing its variability as depicted in the variation map (see Fig. 8). Furthermore, the variability (as expressed in number of variants) and number of sub-processes are as most intensive in the middle section of the main process as can be seen in Fig. 12. For "approve", "match", and "settle", there are three levels of decomposition.

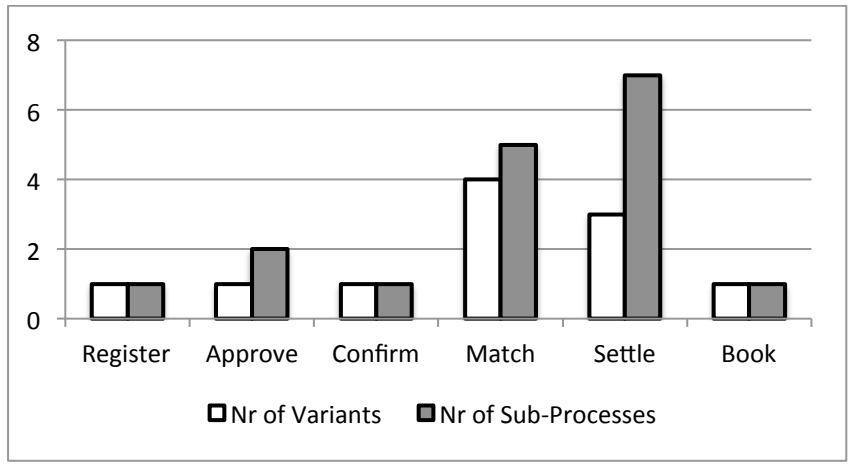

Figure 12. Number of variants and sub-processes in the main process

### 6.2 Results of the DNA Sequencing Case Study

We recall that in the DNA case study, we modelled the same set of processes using the decomposition-driven method and the S&M method as a baseline. The baseline method led to 110 process models, comprising 379 activity nodes (only counting sub-processes and tasks). The duplicity count is 218 (as defined in Section 6.1). Thus the process models in the baseline method had a duplication rate of 41%. On the other hand, the process variants modelled according to our method had 379 activity nodes with a duplicity count of 92 (i.e. 20 %). Compared to the baseline, our method resulted in a reduction of duplication of about 50%. Furthermore, our method led to 83 (sub-)process models whereas the baseline required 110, i.e. 33% more. These observations (summarized in Table 4) support the hypothesis that by applying the decomposition-driven method, a family of process variants can be represented using fewer activities and sub-process models compared to a fragmented approach.

**Table 4.** Size metrics for our method versus baseline scenario.

| Variable | S&M | Decomposition-driven method |
|---|---|---|
| Process Models | 110 | 83 |
| Sub-Processes | 147 | 93 |
| Activity Nodes (Tasks) | 379 | 371 |
| Duplicate Activity Occurrences | 218 | 92 |
| Duplication rate | 41% | 20 % |
| CNC | 0,9 | 0,97 |

Similar to the first case study, we used the CNC metric to compare the output of our method with the baseline. In the baseline scenario, each variant of for instance "prepare sample" was modelled separately. This resulted in a total of 7 sub-processes, one for each variant of "prepare trueseq sample". These models have a CNC of between 0.88 and 0.91. However, as they were similar and lacked strong variation driver, they have been modelled together. As such, the complexity of this sub-process is higher as it encompasses a total of 7 variants and has a CNC of 1.24. We note, as in our previous case, a trade-off between number of process models and complexity. The baseline process models have a total of 739 arcs and 822 nodes (sub-processes, tasks, gateways and events), which gives an average CNC of circa 0.9.[2] On the other hand, the set of process models obtained via our method have 753 arcs and 773 nodes, thus an average CNC of around 0.97. This marginally higher CNC value should be contrasted with the significant reduction in the number of process models (110 versus 83) and duplication rate (41% versus 20%).

Similar to the FX and MM case, we note that the variability in the process occurs more towards the mid section of the main process. In this case, there are four sub-processes of the main process as can be seen in Fig. 9. The first sub-process, "register sample project" does not have any variability on the level of the variation map. However, at lower levels of decomposition, there is variability in terms of method used to measure the quality of the samples. In "prepare sample" there are five variants at the level of the variation map and in total, within this sub-process, 15 variants. In "process sample" there are a total of 8 variants and finally, in "deliver sample data" there is only one variant. In total, the DNA case has 27 variants. This is also reflected in the number of sub-processes under each main sub-process of the main process. In "register sample project" there are a total of 6 sub-processes. The numbers for "prepare sample", "process sample" and "deliver sample" are 64, 22 and 1 respectively. The DNA case has more levels of decomposition (up to five levels) as compared to the baseline scenario (3 levels). As such, the baseline scenario has fewer levels of process model hierarchy (more flat) but is larger whereas the models discovered through our method, have more levels of hierarchy (more deep) but overall smaller.

### 6.3 Threats to Validity

When conducting case studies, there are threats to validity that ought to be considered, particularly regarding construct validity, external validity and reliability [17].

External validity concerns the extent to which the findings can be generalised beyond the setting of the study. Our method has been applied on two case studies, and in line with the inherent limitation of case study methods, the results are limited in the extent they can be generalised. The results are naturally dependent on the domain experts and the purpose of the study, and it can, therefore, limit the repetitiveness of our decomposition driven method. Hence the method is replicable but results may vary due to aforementioned reasons. In addition, the FXMM case was perhaps easier to manage due to relatively less number of process models and in the DNA case was not of high degree of complexity (in terms of having mostly sequential processes). As such, the application of the method has not been tried and must be seen as a limitation on the generalising of the results for the time being. On the other hand, it should be underscored that the case studies have been conducted in two different industrial settings and contexts with active involvement of domain experts with very different backgrounds.

---

[2] This relatively low CNC value stems from the fact that a majority of processes are derived from laboratory protocols that are highly sequential (i.e. relatively few branching points).

Reliability refers to the level of dependency between the results and the researcher, i.e. would the same results be produced if another researcher conducted the study. This threat was to some extent tackled by having several series of verifications by the domain experts without the presence of the researcher (member checking as defined in [24]). Furthermore, we applied peer debriefing [24] to further ensure better reliability. In addition, by applying both data triangulation (using both protocols and domain experts) and observer triangulation (two authors of this paper involved), we reduced the threat to validity [24].

Construct validity refers to the extent to which the tools used measure what the researcher has in mind and also what is being investigated. The risk construct validity was reduced as the results of the case studies, were de facto used. The consolidated process models for the banking case study were used in a four-day workshop with a supplier of off-the-shelf solution to investigate the extent to which the solution could satisfy their needs. The discovered process models for the DNA case study were used to derive requirements for their future LIMS.

## 7   Related Work

The contribution of this paper related to three areas of business process modelling: (i) variability modelling in business processes, (ii) standardization and consolidation of business processes, and (iii) discovery of business processes with variants. Below we review related work in these three areas.

### 7.1   Variability Modelling

Over the past decade, a number of approaches have been proposed to model families of process variants. The common aim of these approaches has been to provide a means to represent a family of process variants in a consolidated manner (i.e. a single model) from which each variant of a process can be derived by application of certain allowed transformations. The main commonality of all such approaches such as C-EPC [25], C-iEPC [26], PESOA [27], and PROVOP [5] is that they are based on annotating various elements of the process models and thereby captures variability. Their strength lies in these annotations combined with the constraints over alterna-tives of allowed and restricted combinations of the process model. Our method can be used in conjunction with these notations given that they encompass the notion of configurable gateways. We have illustrated our method with plain BPMN but our method is in no way prescriptive in terms notational language to be used. In fact, if a configurable process model language were to be used, it could be more specific in, for instance, capturing variability in the variation map, as their notation can capture more semantics as compared to plain BPMN. However, we provide a method for eliciting the relevant information needed to model families of process variants where both syntactic and business reasons for variability are considered. As such, our contribution is complementary to the above-mentioned variability modelling approaches.

In [28], an approach to manage flexibility in process models when dealing with both design time and with run-time situations. The main idea of this approach is that a core process (i.e. main process) consists of identifiable and pre-defined activities, which can contain pockets of flexibility. These pockets of flexibility can be seen as separate sub-processes that contain activities (or further sub-processes) together with rules defining allowed sequence of execution of activities and constraints. Our work is distinguished in two ways. Firstly, we limit our method to design time variability whereas this approach [28] stresses the run-time flexibility. Secondly, our method provides a method for consolidating or discovering families of process variants that can be represented with plain BPMN or according to other approaches such as pockets of flexibility.

Our work is also related to variability modelling in software product lines (SPL) where variability modelling is predominantly captured by feature models. These approaches have been extensively studied [29] and one such example is proposed by Schaefer [30]. In this paper, an approach for model driven development of software intensive systems is proposed that begins with a core model (feature diagram) that captures a valid product. Then, each level of refinement, by extension of the core model, layer of models is added that specify applicable features configurations. As such, applying additions, modifications and/or removal of model fragments of the core model, creates the layers of models under the core model. As these approaches are based on feature models, they take the viewpoint of the product and are primarily aimed at describing product variations in a static way whereas in a process model, the focus is on how such an instance (feature) is produced. Furthermore, our method is applied in consolidation and discovery of families of variants, which is not the primary focus of SPL domain. However, it should be noted that our method transposes ideas behind feature diagrams to the field of process modelling. Indeed, variation matrices and variation maps can be seen as integrated views of process models and the features that drive variations in the underlying processes.

### 7.2 Standardization and Consolidation of Processes

*Process standardisation* is the act of merging multiple variants of a process into one standard process [31]. This is different from process model consolidation, which instead seeks to merge multiple process models (not the processes themselves) into a single model. One step in process standardisation is to identify suitable processes that can be standardised. Proposed methods to achieve this include assessing process complexity (different from model complexity) [32] or applying user-centred design approaches such as work practice design, to aid with identifying candidate processes based on how employees perform their responsibilities [33]. Since our method focuses on model consolidation and not process standardisation, it does not touch upon the organizational change management issues that are central in standardisation. This having been said, process model consolidation and process standardisation share common concerns. In particular, we foresee that the business variation drivers identified via our method could serve as input for standardisation decisions.

Related to process standardisation is *process harmonisation*, which seeks to reduce the differences between models of variants of a process [31] rather than aiming at merging these models. Romero *et.al.* [34] propose a technique to determine an optimal level of process harmonisation based on the identification of so-called influencing factors (i.e. variation drivers) and based on similarity metrics between the models of the individual variants. Their method, however, requires that the process models would be represented at a low level of details. In contrast, our method can be applied when the process variants are not modelled at the same level of detail or when the models are incomplete (e.g., some business processes have not been modelled or not modelled at the same level or using the same conventions as others).

Alternative methods to process model consolidation include automated process model merging methods such as the one proposed by La Rosa et.al [4]. In these methods, multiple variants of a process model are merged into a single model, essentially by identifying duplicate fragments and representing these fragments only once in the merged model. This and similar approaches such as approximate clone detection[35] have the limitation of being based purely on syntactic similarities across process models. They do not take into account business drivers. Our method can be seen as an approach to answer the question of where it makes sense to merge, and where it is better to keep separate models. Specifically, given as input a set of models of process variants, structured as a hierarchy, we can apply our method to identify sub-processes belonging to different variants that could be merged, using for example the automated method of La Rosa et al. [4] so as to decrease the size of the models. In the case studies considered in this article, it was not possible to apply automated process merging in this way, because there was no input process hierarchy. In the FX/MM case study, the input models were flat, whereas in the DNA case study there were no input models. And automated merging of the four flat FX/MM input process models, would only lead to a much larger flat models.

Finally our work is related to process model refactoring [36], where the aim is to rewrite process models in order to improve their comprehensibility, maintainability or reusability, but without altering their execution semantics. Weber *et.al.* [37] propose a catalogue of "smells" in process models that could be treated as candidates for refactoring. One such "smell" occurs when similar fragments appear repeatedly in the same process model or across multiple models in a collection. This duplication can be tackled by extracting similar fragments into shared sub-processes (also known as shared sub-process extraction). Shared sub-process extraction is complementary to our method, insofar as we can apply it to the output of our method in order to further reduce size and duplication. It does not however replace our method in that sub-process extraction does not tell us which parts of a family of processes should be modeled together versus separately.

### 7.3 Discovery of Business Processes with Variants

Our second case study falls under the domain of process model discovery. Methods for process model discovery can be broadly classified into automated ones and manual ones. Automated methods exploit existing data to generate a process model. In this line, one sub-category of approaches is concerned with the discovery of process models from execution logs, also called "process mining" [38] or "workflow mining" [39,40]. Some of these approaches use trace clustering to uncover potential variants of a process, arguing that variants would manifest themselves as clusters of similar traces in the logs. In other words, they focus on syntactic similarity as the basis for variant identification, putting aside business drivers for variation, which are not identifiable from execution logs.

Another sub-category of approaches for automated process discovery is those that take as starting point textual documentation. For example, Ghose *et.al.* [41] propose a framework for Rapid Business Process Discovery (R-BPD). Their framework is based on querying text artefacts such as corporate documentation to create initial process models that are subsequently edited by domain experts. Again, their method manages variations in process models from a syntactical perspective and does not consider business drivers, nor does it consider the possibility of alternating discovery of variants with process decomposition.

Non-automated process model discovery methods are concerned with collecting, organizing and analysing data from various stakeholders in order to produce process models. The method by Sharp and McDermott [9] – which we used as a baseline – is an exemplar of a method in this field. Another similar method is presented in [42].

A related field is that of process architecture, which is concerned with identifying and organising processes of an enterprise [43]. Process decomposition is one aspect of process architecture. In this field, sets of guidelines have been proposed to achieve a decomposition that reduces complexity [43–47]. However, these guidelines do not state how to identify process variants and are thus complementary to our work.

The method for process identification defined in [48] is also closely related to ours. This latter method begins with the identification of cases (variants) and functions that should be included in the process architecture. Next, a case/function matrix is created and by applying a set of 8 guidelines, processes are identified from this matrix. Two steps in this method explicitly deal with variations. In the first of these steps, variants of a business process (called cases) are listed. Later, in a second relevant step, if a process model for one case is found to be syntactically very different from the model of another variant, the two variants are explicitly separated. However, the method in [48] does not consider business drivers for variation, nor does it consider the possibility of alternating process decomposition with variant identification.

# 8 Conclusion

This paper addressed the question of how to manage the trade-off between modelling multiple variants of a business process together versus modelling them separately. In this context, we investigated the following propositions: (1) Decisions to model multiple process variants separately should be taken at the lowest possible level of process decomposition rather than upfront. In other words, rather than deciding upfront to split multiple process variants into separate models, one should consider postponing this decision at the level of each sub-process, until there are strong reasons for modelling the variants separately. (2) Decisions on whether to model variants together or separately should be based both on the business drivers for variation (the extent to which the separation between variants are integral to the business) as well as syntactic drivers (the degree of similarity between variants).

Based on these propositions, we presented a decomposition-driven method for modelling families of process variants. We validated this method by applying it to two case studies: one aimed at consolidating an existing collection of models, and another aimed at modelling of a family of process variants from scratch.

Although not fully generalizable, the case study findings show that the proposed method provides a structured approach to modelling families of process variants in a way that reduces duplication in the resulting process models, with relatively minor penalty on model complexity.

The method has been formulated in an intra-enterprise setting where all the stakeholders are able to agree on the primary drivers of variability in the business processes. When applying the method in a cross-organisational process, additional issues might arise. For example, two business partners might have different viewpoints of the relative strengths of the drivers. This situation would require a compromise, which is not considered in our method. Furthermore, the method has been developed in the context of a procedural process modelling language (e.g. BPMN) where decision points are explicitly represented along the flow of activities. Recently, alternative process modelling paradigms based on declarative styles have been proposed [49]. In declarative process models, activity flows and decision points are neither exhaustively nor explicitly captured and hence the proposed method is not directly applicable. Extending the method to cater for cross-organisational processes and declarative process modelling are avenues for future work.

In addition, during the execution of the second case study, we found anecdotal evidence that data objects plays a key role in determining how process variants differ from one another. Specifically, we identified situations where two variants of a sub-process were highly similar at the procedural level, but differed significantly in terms of the input objects. In some cases, we found even that the same atomic task (e.g. prepare consumables) would differ across two variants because different chemicals would be employed and their preparation would require different steps. This finding suggests that object variability can also affect the decision to model variants together or separately. In future, we plan to investigate the interplay between business process variability and object variability, in view of designing an integrated method that bridges these two sides of the variability equation.

**Acknowledgement:** This research was supported by the European Social Fund via the Doctoral Studies and Internationalisation Programme – DoRa.